\documentclass[twocolumn,aps,pra,showpacs,showkeys,preprintnumbers,amsmath,amssymb]{revtex4-1}
\usepackage{graphicx}
\usepackage{dcolumn}
\usepackage{bm}
\usepackage{epsfig,wrapfig}
\usepackage{epstopdf} 
\usepackage{fancyhdr}
\usepackage{indentfirst}
\usepackage{titlesec}
\usepackage{amsmath}
\begin{document}
\title{Sound Reception System by an Acoustic Luneburg Lens}
\author{Sang-Hoon  \surname{Kim}$^{a}$}
\email{shkim@mmu.ac.kr }
\author{Byeong-Won  \surname{Ahn}$^{a}$}
\author{ Kyung-Min \surname{Park}$^{b}$}
\author{ Gung Su \surname{Lim}$^{c}$}
\affiliation{
$^{a}$ Division of Marine Engineering, Mokpo National Maritime University, Mokpo 58628, R. O. Korea
\\
$^{b}$ Division of Marine Engineering $\&$ Coast Guard, Mokpo National Maritime University, Mokpo 58628, R. O. Korea
\\
$^{c}$ Division of Cadet Training, Mokpo National Maritime University, Mokpo 58628, R. O. Korea
}
\date{\today}
\begin{abstract}
Applying the directionality of a Luneburg lens, a sound reception system for ships is designed.
A two-dimensional acoustic Luneburg lens which is made of 253 acrylic pipes of different radii which control the refractive index is used.
It has a large disk shape; a diameter of 180 cm and a height of 20 cm.
The 8 microphones were installed around the edges of the lens.
The sound reception ability of the lens was tested on the sea using the horn
of a university training ship as an emitter and the lens loaded on a  small boat as a receiver.
Both are moving at 20 $\sim$ 40 km/h at a distance of between 300  $\sim$ 700 m.
The lens could focus to recognize the direction of the input sound in the frequency range
between 300  $\sim$ 900 Hz.
\end{abstract}
\keywords{Acoustical measurements and instrumentation, Acoustical lenses and microscopes,
Acoustic sensing and acquisition}
\pacs{43.58.-e, 43.58.Ls, 43.60.Vx}
\maketitle
\section{Introduction}

\begin{figure}
\resizebox{!}{0.14\textheight}{\includegraphics{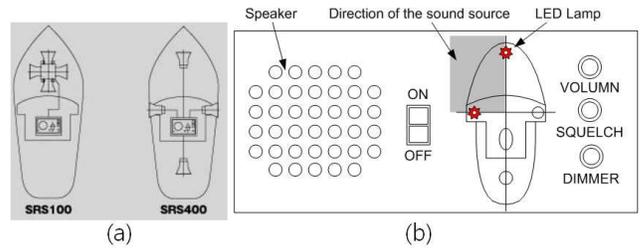}}
 \caption{(a) Positions of the SRS100 and SRS400 microphone units.
 (b) Panel in the ship's bridge to show how to work the commercialized SRS.
 The second quadrant or the shaded area is the direction of the sound source.}
  \label{srs}
\end{figure}

Many ships are heavy and therefore it's not easy for a ship to turn or stop like an automobile
due to large inertia.
Every ship is equipped with radar which gives the information of other ships nearby.
Nevertheless, there is a possibility of causing an accident due to poor visibility
such as from dense fog or heavy rain.
For this reason, all vessels shall be subject to short sound (1 s)
and long sound (4  $\sim$ 6 s) to prevent a collision accident by attaching a sound generating device by IMO (International Maritime Organization) regulation \cite{imo}.
As a device to support this, SRS (Sound Reception System) in Fig.~\ref{srs} has been developed and used for ships \cite{srs}.
The SRS is composed of one speaker and four lamps.
Four weatherproof microphones are mounted on the front, tail,
and both sides of the ship and connected to the device.
If two lamps are lit with speaker beeps,
then the quadrant area of the two lamps is the direction of the sound.
It provides one of four possibilities to the navigator as to the location of the acoustic source.

In this paper, we suggest a totally different SRS by a metalens so called acoustic Luneburg lens (ALL).
Luneburg lens is a GRIN (GRadient INdex) lens and focuses the incoming wave on the opposite side of the lens without aberration \cite{lu,gu,mo}.
There has been a lot of progress in the realization of the Luneburg lens
 due to metamaterials \cite{ch,fa,ze}.
For example, the focusing area could be any point, and the shape does not necessarily
have to be spherically symmetric \cite{ku}.

   In 2014 Kim et al.\cite{kim,kim1,kim2} reported the realization of ALL.
    Afterwards, several ALL were presented due to many possible application areas
    \cite{chen,fang,cum1,cum2}.
ALL has two serious properties and several potential application areas related to it.
One is the focusing ability that is free of aberration and the other does the directional ability from the other side focusing.
The focusing ability is useful for ultrasonic diagnosis in the biomedical or industrial fields because it extends the sensing range.
The directional ability is maybe useful for sonar because it gives the direction of the acoustic source directly.
It works passively by the focusing of a sound wave, and actively by creating a plane wave at both modes.
 However, a direct application of ALL in industry has not been reported yet.
We fabricated a two-dimensional (2D) ALL as SRS for ships and tested the directional ability on the sea.
The result of the field experiment is reported.

\section{Design and fabrication}

\begin{figure}
\resizebox{!}{0.32\textwidth}{\includegraphics{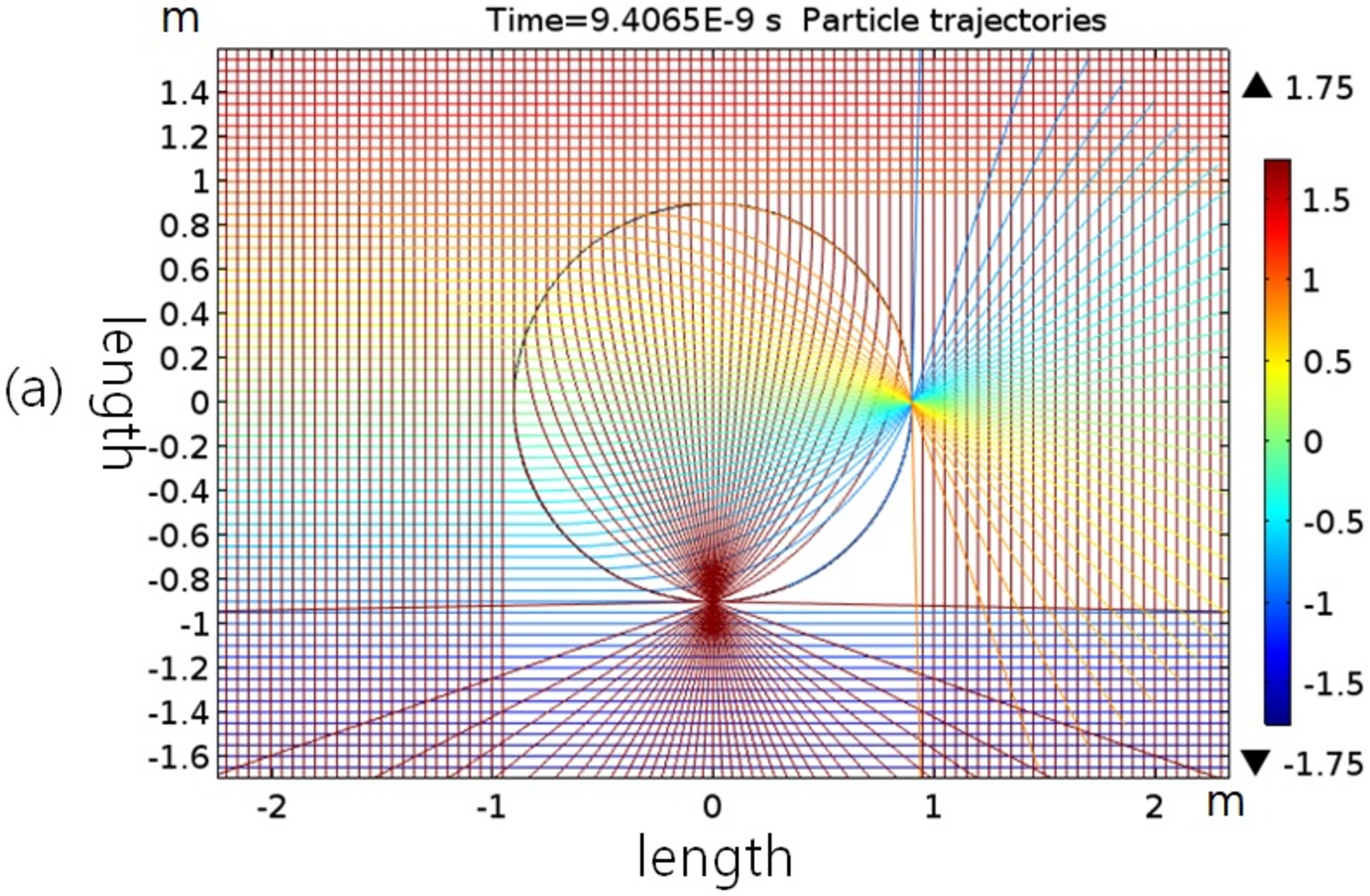}}
\resizebox{!}{0.33\textwidth}{\includegraphics{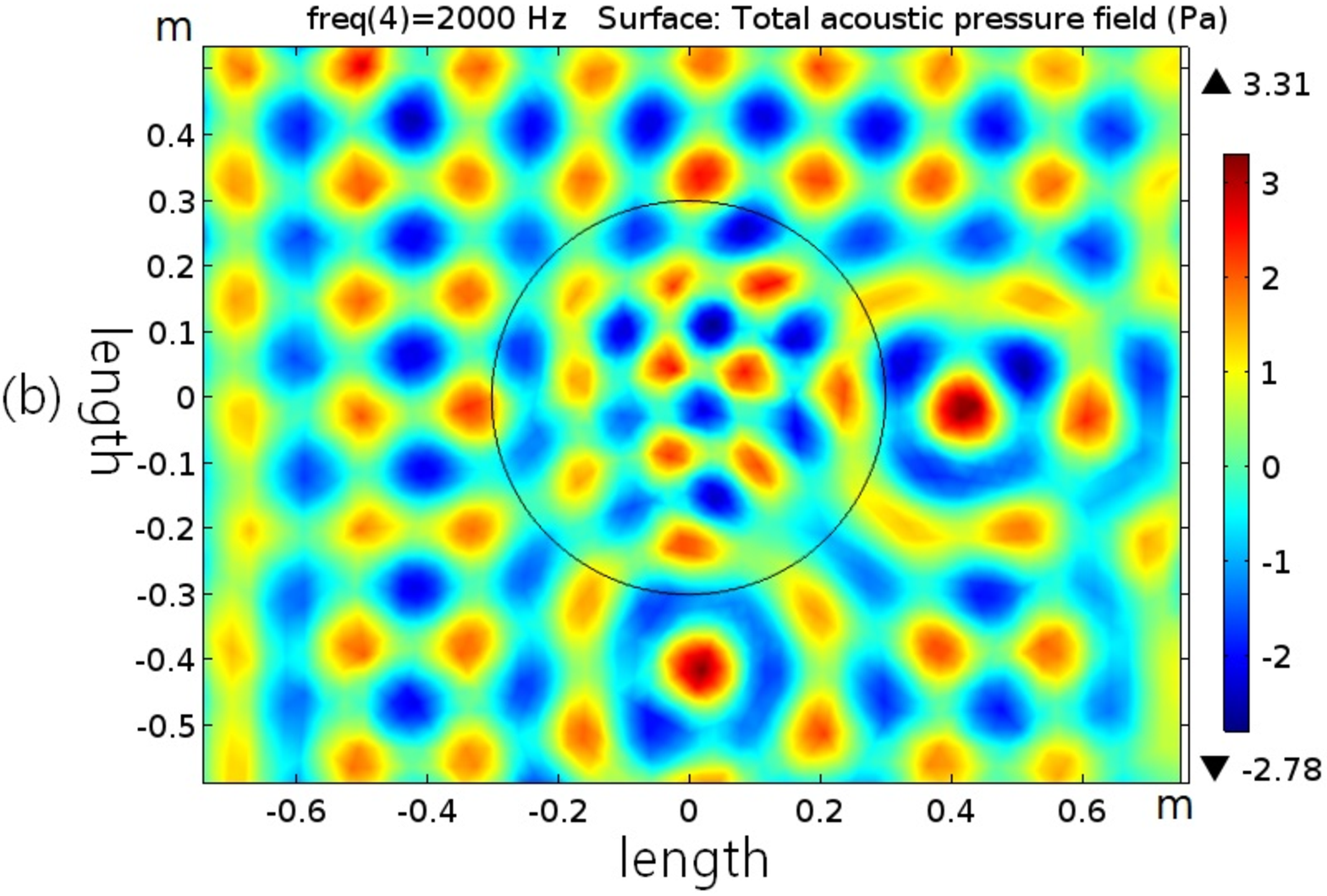}}
 \caption{Numerical simulation of multifocusing by Luneburg lenses.
 (a) An optical and (b) an acoustic Luneburg lens.
}
  \label{multi}
\end{figure}

A main feature of the Luneburg lens is its multifocusing property,
as it is shown in Fig.~\ref{multi}(a) and Fig.~\ref{multi}(b) for the optical and the acoustical versions, respectively.¡±
   We designed and fabricated a 2D ALL using the common GRIN lens method.
It is a direct geometric application of transformation optics.
The refractive index profile of the Luneburg lens is given as
$ n(r)=\sqrt{2- (r/R)^2}, $
where $R$ is the radius of the lens and $0 \leq r \leq R$.
It was derived from Fermat's principle and the calculus of variation \cite{lu,gu,mo}.
The refractive index can be rewritten in the discrete form too as
$ n_i = \sqrt{2-(i/N)^2}, $
where $N$ is the number of layers inside the lens and $i=0,1, 2, ... ,N-1$.

\begin{figure}
\resizebox{!}{0.25\textheight}{\includegraphics{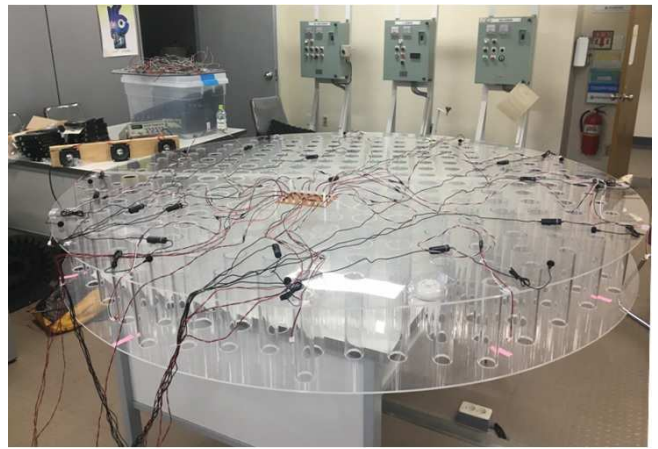}}
\caption{The fabricated two-dimensional
acoustic Luneburg lens made of 253 acrylic pipes of different radii.}
\label{all}
\end{figure}

The velocity of the sound waves traveling inside the ALL depends on the density and bulk modulus.
Then, the effective density in the medium is the only parameter controlling the acoustic refractive index locally \cite{jose1}.
    The lens is composed of 253 square cells and each cell has a high impedance cylindrical obstacle  in the centre.
    We set the side length of the square cell be 10 cm and the number of concentric rings is $N=10$.
Then, the diameter of the lens is 180 cm.
The obstacles are acrylic pipes of different radii and the same height of 20 cm.
The lens of the top and bottom was covered by two large acrylic plates to satisfy 2D condition.
 The thickness plates and the pipes were 5 mm.
The lens fabricated finally is shown in Fig.~\ref{all}

\begin{figure}
\resizebox{!}{0.3\textwidth}{\includegraphics{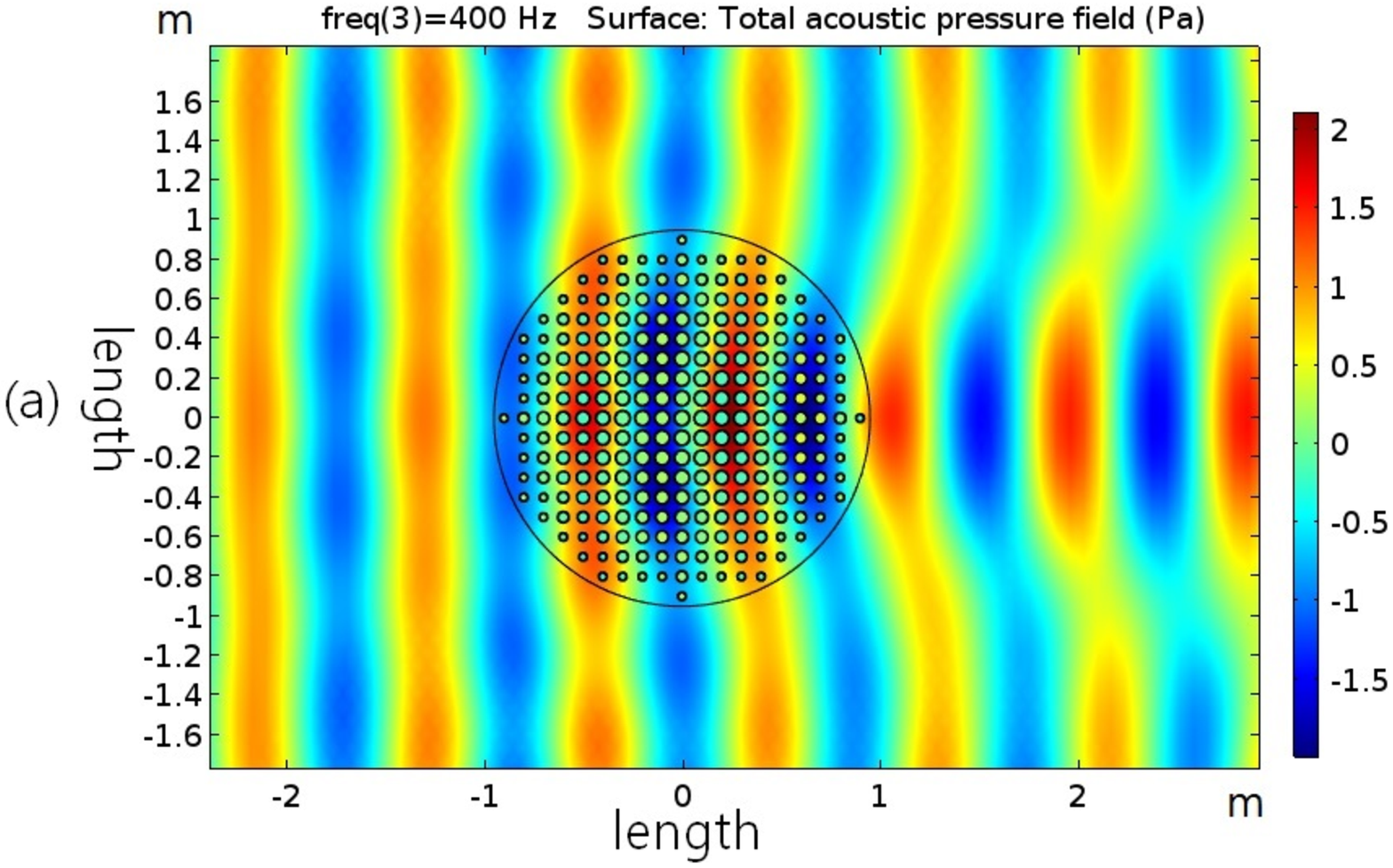}}
\resizebox{!}{0.3\textwidth}{\includegraphics{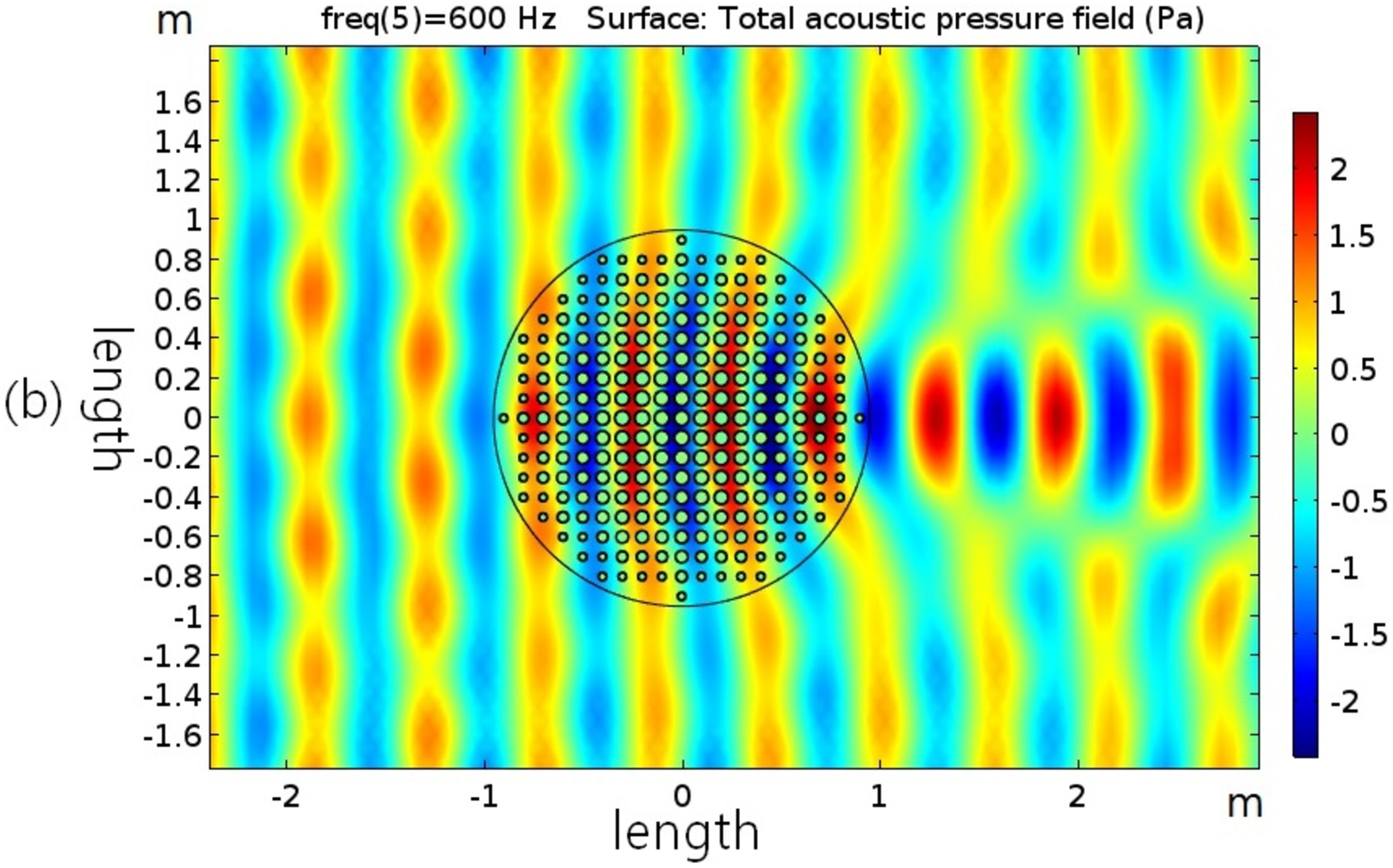}}
\resizebox{!}{0.3\textwidth}{\includegraphics{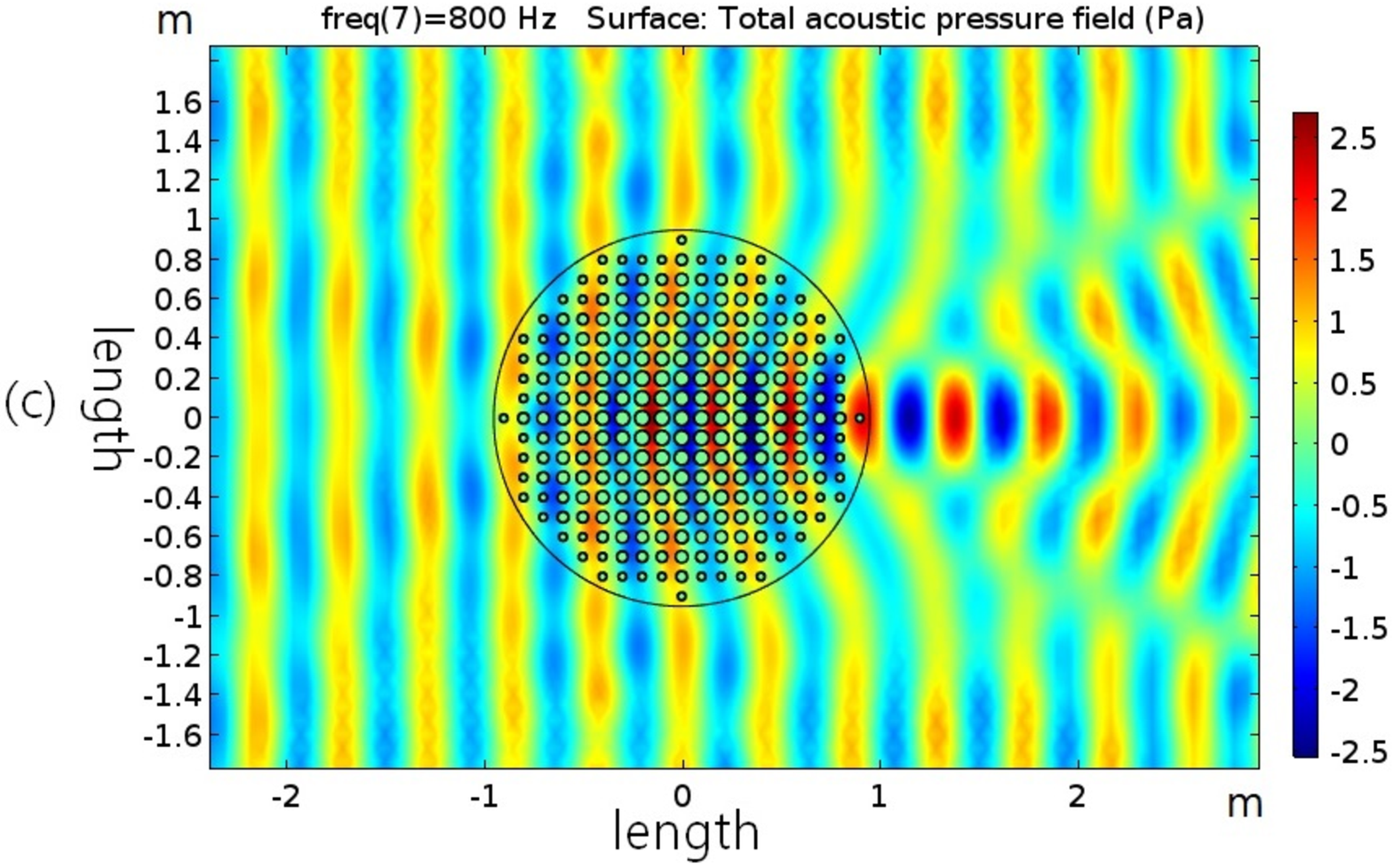}}
\resizebox{!}{0.32\textwidth}{\includegraphics{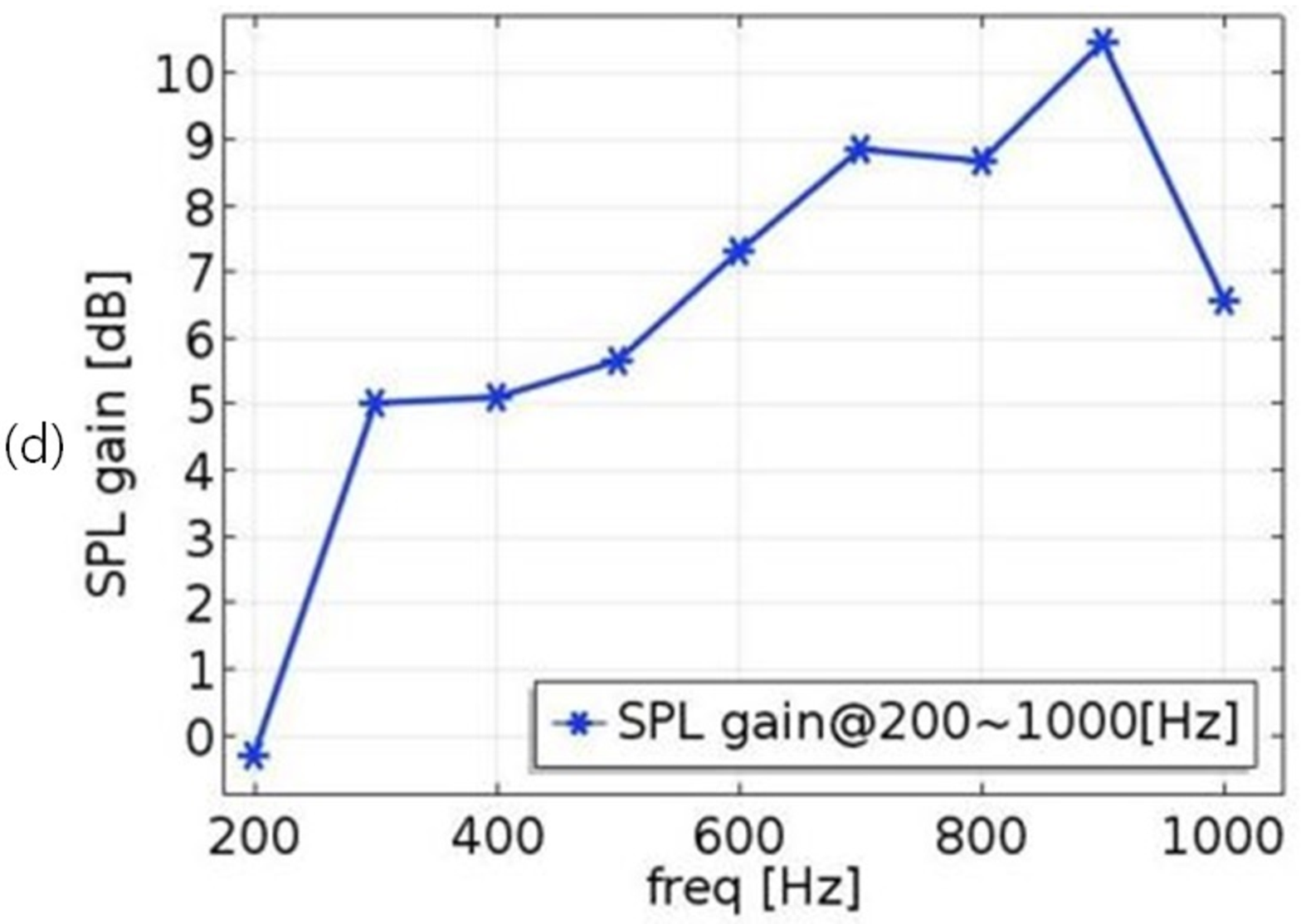}}
\caption{Numerical simulation of the acoustic Luneburg lens
at different frequencies and their sound pressure level.
(a) Freq.= 400 Hz, (b) Freq.= 600 Hz, (c) Freq.= 800 Hz.
(d) Frequency dependent acoustic gain.
 }
\label{simul}
\end{figure}

The operational wavelengths of the lens are in the range $a < \lambda < 2R$,
where a is the size of the unit cell and R is the radius of the lens.
  At the same time the thickness of the lens or the height of the pipes, 20 cm,
must be smaller than the wavelength due to the 2D restriction.
Then, the theoretical maximum focusing frequency range is about 200 Hz $< f <$ 1,700 Hz.
However, the actual available frequency range is much narrower than this.
We plotted a numerical simulation of ALL at some frequency ranges by COMSOL Multiphysics in Fig.~\ref{simul}.
The SPL(sound pressure level) gain of 5$\sim$10 dB was obtained
in the available frequency range of 300 Hz $< f <$ 900 Hz.

\section{Experiment on the sea}

We have tested the focusing ability of the ALL with a ship's horn sound a couple of years ago \cite{kim3}.
The ship is docked and ALL was fixed on land.
  Then, the lens focused the incoming sound enough to recognize up to 1.4 km away from the source   without any other noise.
 At the distance of more than 1 km, lower than 200 Hz was the main source from the horn.

\begin{figure}[h]
  \centering
\resizebox{!}{0.13\textheight}{\includegraphics{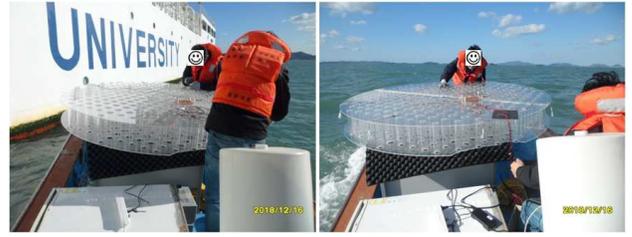}}
  \caption{Installing the microphones to the ALL on the boat.
The edge of the ALL was broken a little during the loading. }
  \label{exp1}
\end{figure}

    However, the real condition at sea is totally different from this.
Sound sources are possibly multiple and emitters and receiver are both moving.
It is pretty dynamic and confusing with a lot of noises.
 Therefore, we decided to test the focusing ability of the ALL at sea where both are moving.
The air horn of the university training ship of MMU (Mokpo Maritime University),
about 4,700 tons, was used as a sound source.
The lens and four people boarded a small boat to control the lens as Fig~\ref{exp1}.
During the loading of the lens to the boat, the acrylic edge of the ALL was broken a little.
  Eight microphones were installed at the edge of the lens with the same intervals of 45 degrees.
Eight is the minimum number of the microphones to check the operation,
and it is adjustable to as many as one needs.
The microphones are connected to the laptop and operated by LabVIEW program.

\begin{figure}
 \resizebox{!}{0.45\textwidth}{\includegraphics{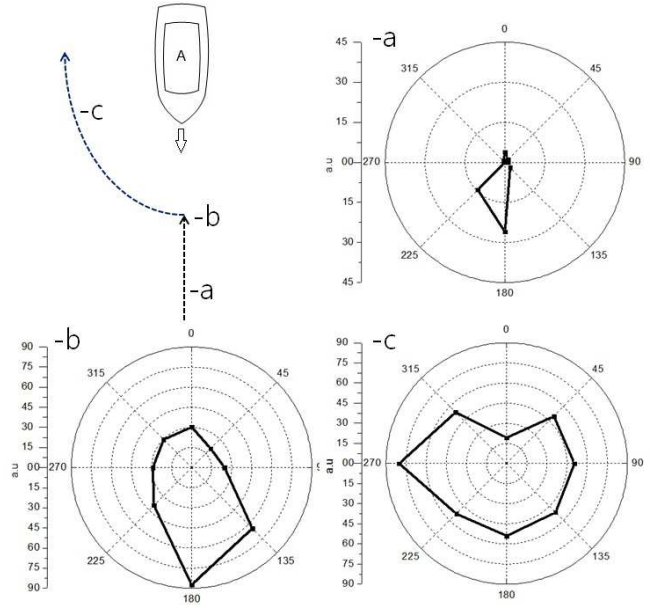}}
  \caption{The trajectory of the two vessels and the data depending on the positions.
A is the emitter and a, b, c are the receiver positions. This is not the real scale.
The distances between the two vessels are
   (a) $\sim$ 700 m, (b) $\sim$ 300 m, and  (c) $\sim$ 300 m.
The round points of 1-8 are the directions.
The bar numbers are the intensities of arbitrary units.
}
\label{nearfar}
\end{figure}

The experiment was implemented on the Yellow Sea
 near 34.47.0N and 126.15.0E on the morning on 16th of December, 2018.
The sky was clear and the temperature was about 3 $\sim$ 5 Celsius.
The training ship and boat were sailed at 20 $\sim$ 40 knots (37 $\sim $54 km/h),
and the distance between the two were about 300 $\sim$ 700 m.
Two vessels moved a similar way of collision situation as Fig.~\ref{nearfar}.
The distance of more than 1 km was too weak to recognize the source signal with the lens.
To detect far distances, it needed a wider lens.

Since we need low frequencies to predict the exact direction, we filtered the high frequency noises of the sound source.
Several MEMS (Micro-Electro Mechanical Systems) microphones ADMP401 were used to convert sound signal into an electrical signal.
It has a SNR (Signal-to-Noise Ratio) of 62 dBA and flat wide-band frequency response from 100 Hz to 15 kHz.
Amplified signal by the amplifier OP37 was transmitted to NI CompactRIO-9022 with NI-9215 which is an analog input module.
FPGA (Filed Programmable Gate Array) design for fast operation was applied.
Input signal is filtered with appropriate data for size comparison at each position by NI-LabVIEW software program.

We did not connect lamps to the lens during the sea experiment.
However, if it is commercialized,
 a monitoring system of the mirror image can be installed in the navigation room.
Then, the motion of the acoustic source can be observed directly by the naked eye and ear.
As approaching the source, the lamps become brighter, and as receding the source,
the lamps become dimmer \cite{video}.
The volume of the speaker works on the same principle.
For those Radar-like SRS a better filtering system in the navigation room is needed.

\section{Discussion}

A two-dimensional acoustic Luneburg lens with a diameter of 180 cm was used as a sound reception system for a ship.
The focusing ability was measured under dynamic condition that the source and receiver were moving at sea.
It directly indicated the direction of sound sources with much better accuracy than the commercialized SRS devices on the market.
  Transparent acrylic was chosen as the material to show the structure of the lens,
 but weatherproof aluminum will be a good choice for manufacturing the copies in the factory using a mould.
The design and performance of the lens as a sound reception system  was reported.
 We tested the performance of the ALL in the air, but it should work underwater as well.
 Therefore, it may be used as a next generation of sonar or underwater acoustic window.

 Although the ALL worked as a SRS in the frequency range between 300 $\sim$ 900 Hz,
it should be at least 5 m of the diameter to be commercialized
because IMO requires to accept the frequency range between 70 $\sim$ 820 Hz \cite{imo}.
Therefore, the circular ALL is suitable for a large ship only.
For small ships, a modification of the lens should be introduced, by making it with elliptical or box-type shape \cite{dyke,hao}.
Another practical concept is foldable lens like wings of insects that expand only during bad weather conditions.

\acknowledgements
The author, S.-H. Kim, thanks to Prof. J. S$\acute{a}$nchez-Dehesa for fruitful discussions.
 

\begin{thebibliography}{99}

\bibitem{imo} IMO resolution MSC.86(70),
``Adoption of new and amended performance standards for navigational equipment (Dec. 1998),
SOLAS Chapter V. Safety of Navigation, REGULATION 19,
``Carriage requirements for ship borne navigational systems and equipment."
ISO 14859:2012 - Ships and marine technology - sound reception systems.
 https://www.iso.org/obp/ui/{\#}iso:std:iso:14859:ed-1:v1:en.

\bibitem{srs}
https://www.phontech.net/product-category/sound-reception/
http://www.ehanshin.co.kr/bbs/sub0208/250421
https://cirspb.ru/en/equipment-and-service/sound-reception/
http://www.omega-in.com/SOUND-RECEPTION-SYSTEM.
https://www.pacatlantic.com/zollner/product-category/sound-reception-systems/
https://www.mackaycomm.com/products
/communication/soundreceptionsystems/zenitel-vingtor-sound-reception-system-vss-2/

\bibitem{lu} R. K. Luneburg, {\em Mathematical Theory of Optics}
(Rhode Island, Brown Univ., 1944).

\bibitem{gu} A. S. Gutman, Modified Luneburg Lens,
 J. Appl. Phys. {\textbf 25}, 855 (1954).

\bibitem{mo} S. P. Morgan,
 General solution of the Luneburg Lens Problem,
 J. Appl. Phys. {\textbf 29}, 1358 (1958).

\bibitem{ch} Q. Cheng, H. F. Ma, and T. J. Cui,
Broadband planar Luneburg lens based on complementary metamaterials,
 Appl. Phys. Lett. {\textbf 95}, 181901 (2009).

\bibitem{fa}  A. D. Falco, S. C. Kehr, and U. Leonhardt,
 Luneburg lens in silicon photonics,
 Opt. Exp. {\textbf 19}, 5156 (2011).

\bibitem{ze}  T. Zentgraf, Y. Liu, M. H. Mikkelson, J. Valentine, and X. Zhang,
 Plasmonic Luneburg and Eaton lens,
 Nat. Nano. {\textbf 6}, 151 (2011).

\bibitem{ku} N. Kundtz and D. R. Smith,
Extreme-angle broadband metamaterial lens,
   Nat. Mat. {\textbf 9}, 129 (2010).

\bibitem{kim} S.-H. Kim, Sound focusing by acoustic Luneburg lens, arXiv:1409.5489v2.

\bibitem{kim1} S.-H. Kim,
Cylindrical Acoustic Luneburg Lens, $8^{th}$ International Congress on Advanced Electromagnetic Materials in Microwaves and Optics (Metamaterials),  364 (2014).

\bibitem{kim2} S.-H. Kim,
Two-dimensional acoustic Luneburg lens, $21^{st}$ Int. Congress on Sound and Vibration, 1 (2014).

\bibitem{chen}
C.-F. Wang, C.-N. Tsai, I-L. Chang, and L.-W. Chen,
Wideband Acoustic Luneburg Lens Based on Gradient Index Phononic Crystal,
Proc. of the ASME 2015 Int. Mech. Eng. Con. and Exp., IMECE2015-52927, (2015).

\bibitem{fang}
R. Zhu, C. Ma, B. Zheng, M. Y. Musa, L. Jing, Y. Yang, H. Wang, S. Dehdashti, N. X. Fang, and H. Chen,
Bifunctional acoustic metamaterial lens designed with coordinate transformation,
Appl. Phys. Lett.,  {\textbf 110}, 113503 (2017).

\bibitem{cum1}
Y. Xie, Y. Fu, Z. Jia, J. Li, C. Shen, Y. Xu, H. Chen, and S. A. Cummer,
Acoustic Imaging with Metamaterial Luneburg Lenses,
Sci. Rep., {\textbf 8}, 16188 (2018).

\bibitem{cum2}
S. Cummer and Y. Xie,
Acoustic Imaging with Metamaterial Luneburg Lens,
J. Acoust. Soc. Am., {\textbf 144}, 1674 (2018).

\bibitem{jose1} D. Torrent and J. S$\acute{a}$nchez-Dehesa,
Acoustic metamaterials for new two-dimensional acoustic devices, New J. Phys., {\textbf 9}, 323 (2007).

\bibitem{kim3} H. Jang, S.-H. Kim, and B.-W. Ahn,
Sound reception system for a ship by an acoustic lens,
J. Korean Soc. of Marine Engineering (in Korean),
{\textbf 42}, 191 (2018).

\bibitem{video} See  https://youtu.be/RVuaeGGwMZY,
 https://youtu.be/jPzrMC7wG3I, https://youtu.be/l45DwmL80Uc
 for the demonstration of the Luneburg sonar in the lab.


\bibitem{dyke} A. Dyke, H. Dyke, S. Haq, and Y. Hao,
Flat Luneburg Lens via Transformation Optics for Directive Antenna Applications,
IEEE Transactions on Antennas and Propagation, {\textbf 62}, 1945 (2014).

\bibitem{hao} J. Gao, C. Wang, K. Zhang, Y. Hao, and Q. Wua,
Beam steering performance of compressed Luneburg lens based on transformation optics,
Results in Physics, {\textbf 9}, 570 (2018).
 \end{thebibliography}
\end{document}